\documentclass[letterpaper,twocolumn,10pt]{article}
\usepackage{usenix2019_v3}
\usepackage{comment}
\usepackage{todonotes}
\usepackage{tikz}
\usepackage{amsmath}

\usepackage[numbers]{natbib}
\newlength{\bibitemsep}
\newlength{\bibparskip}
\let\oldthebibliography\thebibliography
\renewcommand\thebibliography[1]{%
  \oldthebibliography{#1}%
  \setlength{\parskip}{\bibitemsep}%
  \setlength{\itemsep}{\bibparskip}%

}

\usepackage{filecontents}

\begin{document}

\date{}

\title{\Large \bf ReproduceMeGit: A Visualization Tool for Analyzing Reproducibility of Jupyter Notebooks}

\author{
{\rm Sheeba Samuel\href{https://orcid.org/0000-0002-7981-8504}{\includegraphics[scale=0.06]{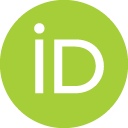}}}\\
Heinz-Nixdorf Chair for Distributed\\
Information Systems\\
and Michael Stifel Center Jena\\
Friedrich Schiller University Jena, Germany\\
{\rm sheeba.samuel@uni-jena.de}
\and
{\rm Birgitta K{\"o}nig-Ries\href{https://orcid.org/0000-0002-2382-9722}{\includegraphics[scale=0.06]{orcid}}}\\
Heinz-Nixdorf Chair for Distributed\\
Information Systems\\
and Michael Stifel Center Jena\\
Friedrich Schiller University Jena, Germany\\
{\rm birgitta.koenig-ries@uni-jena.de}
} 

\maketitle

\begin{abstract}
Computational notebooks have gained widespread adoption among researchers from academia and industry as they support reproducible science.
These notebooks allow users to combine code, text, and visualizations for easy sharing of experiments and results.
They are widely shared in GitHub, which currently has more than 100 million repositories making it the largest host of source code in the world.
Recent reproducibility studies have indicated that there exist good and bad practices in writing these notebooks which can affect their overall reproducibility.
We present \textit{ReproduceMeGit}, a visualization tool for analyzing the reproducibility of Jupyter Notebooks.
This will help repository users and owners to reproduce and directly analyze and assess the reproducibility of any GitHub repository containing Jupyter Notebooks.
The tool provides information on the number of notebooks that were successfully reproducible, those that resulted in exceptions, those with different results from the original notebooks, etc.
Each notebook in the repository along with the provenance information of its execution can also be exported in RDF with the integration of the ProvBook tool.
\end{abstract}

\vspace{-0.2cm}
\section{Introduction}
Several large studies have emerged to analyze different aspects of Jupyter Notebooks, particularly from GitHub.
Rule et al.~\cite{Rule3173606} analyzed over 1 million publicly available notebooks from GitHub.
The focus of their study was on the exploration of the usage and structure of Jupyter notebooks, especially analyzing the use of code, text, and comments inside the notebooks.
Another recent study by Pimental et. al.~\cite{Pimentel2019}, analyzed 1.4 million Jupyter notebooks from the GitHub repositories created between January 1\textsuperscript{st}, 2013 and April 16\textsuperscript{th}, 2018.
They presented a detailed analysis of the quality and reproducibility of these notebooks.
Their focus was not only on the structure of notebooks but also on their execution and replication.
Inspired by these works, we present ReproduceMeGit, an online tool where users can examine any GitHub repository and obtain an extensive analysis of different aspects of notebooks including their structure and reproducibility features.
This tool also provides direct access to Binder~\cite{binder2018} and ProvBook~\cite{ProvBookSamuel}.
Binder is an open-source web service provided by Project Jupyter to create shareable reproducible environments for Jupyter Notebooks in the cloud.
It captures the code repository with its technical environment by creating containers and generates user sessions to run the notebooks in those containers.
ProvBook, an extension of Jupyter Notebooks, captures and visualizes the provenance of the execution of the notebooks.
By using ProvBook in ReproduceMeGit, users can download the notebooks' execution provenance information in RDF.
\vspace{-0.2cm}
\section{ReproduceMeGit: An Overview}
\begin{figure}
\centering
\includegraphics[width=9cm, height=5cm]{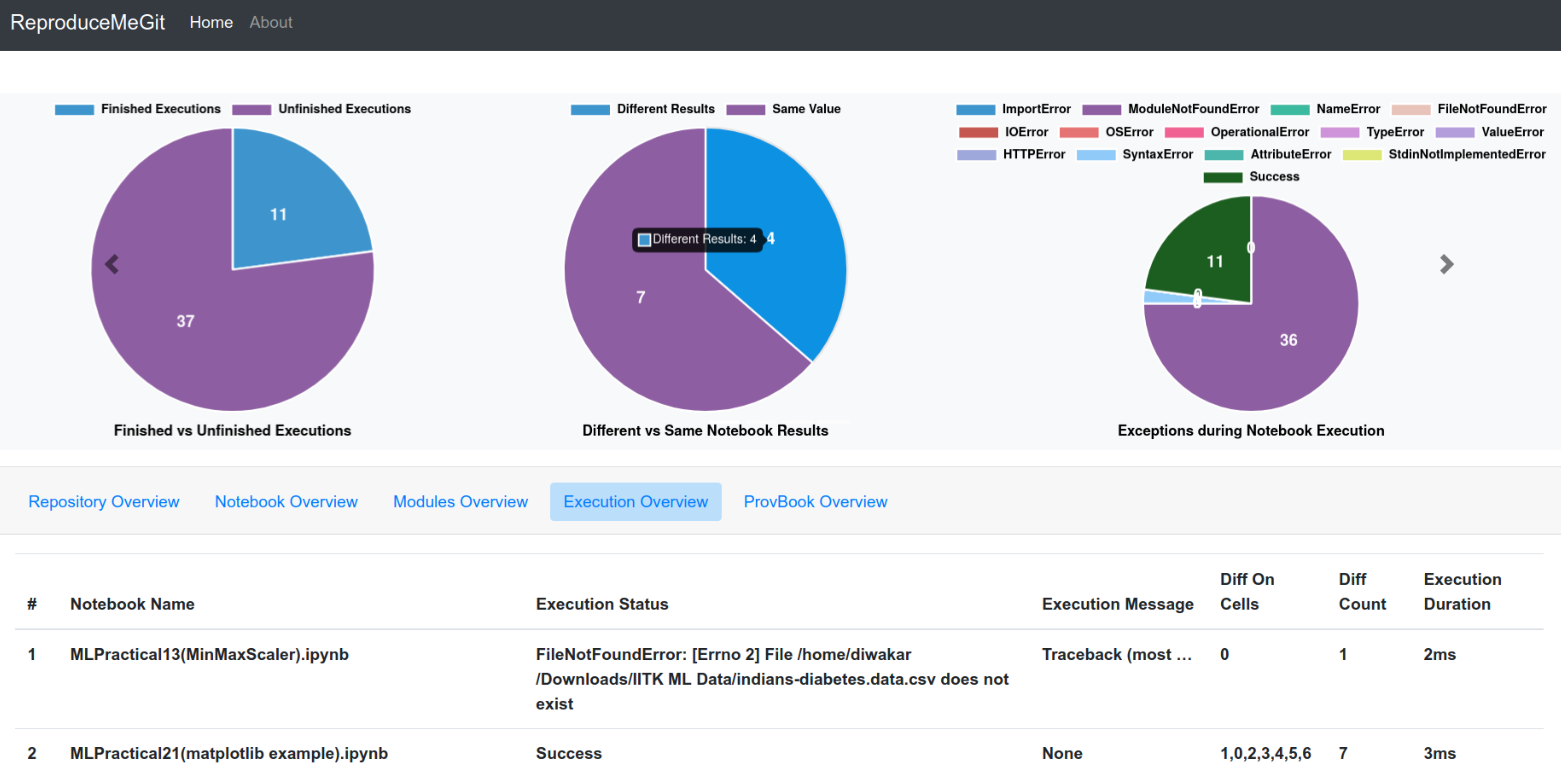}
\caption{Visualizations of the analysis of reproducibility of Jupyter notebooks in ReproduceMeGit}
\label{fig:ReproduceMeGit}
\end{figure}

We present ReproduceMeGit, a visualization tool for analyzing different aspects of Jupyter Notebooks including their structure and reproducibility.
The goal of this tool is to provide an overview of the reproducibility of notebooks in a selected repository by providing information on the number of notebooks that
were successfully reproducible, errors that occurred during runs, difference in the results from the original notebooks, provenance history of runs, etc.
Figure~\ref{fig:ReproduceMeGit} depicts the GUI of the ReproduceMeGit tool.
This tool is built on top of the work from~\cite{Pimentel2019}.
ReproduceMeGit provides a user interface where users can enter a repository URL and reproduce Jupyter Notebooks using a \textit{Reproduce} button.
The tool fetches the content of the repository using the GitHub API, processes the repository and scans it for Jupyter Notebooks.
It then loads the notebooks and extracts information on their structure, cells, modules, etc.
In the next step, to execute these notebooks, the requirements for the environment setup are collected.
The dependencies for the execution of these notebooks are commonly defined in the repositories using  files like \textit{requirements.txt}, \textit{setup.py} or \textit{pipfile}.
All the dependencies are installed using the setup environment files collected from the repository. 
Conda, an open-source package manager and environment management system environment, is used to install and manage the dependencies of Python packages.
After setting up the environment, the notebooks are executed using the Python version of the original notebook.
Supported by ProvBook, ReproduceMeGit stores the provenance information of the execution.
In~\cite{Pimentel2019}, direct string matching is used to calculate the difference between the execution of the original notebook from the repository and their execution.
In contrast, we use the nbdime\footnote{\url{https://github.com/jupyter/nbdime}} tool provided by Project Jupyter to calculate the difference between the two executions.
This is because nbdime provides diffing of notebooks based on the content, especially for image-diffs.

ReproduceMeGit shows the highlights of the reproducibility study in the top panel and the respective detailed analysis in the bottom panel.
It displays the respective count of notebooks  (un-)successfully finishing the executions.
Out of the successfully executed notebooks, it provides information on the number of notebooks that had the same or different results compared to the original.
For notebooks which failed to execute, the tool shows the exceptions that occurred in their executions.
These exceptions include \textit{ImportError}, \textit{ModuleNotFoundError}, \textit{FileNotFoundError}, \textit{IOError}, \textit{SyntaxError}, etc.
It also displays the information of the notebooks which provide output and execution count in their code cells.
The execution count is the cell identifier which denotes the count of the execution of the cell~\cite{sheebathesis}.
It also provides more analysis on the structure of the notebooks.
It shows the number of notebooks that are valid based on their nbformat, kernel specification, and language version.
The distribution of most common programming languages and the version of them used in these notebooks are also shown in the tool.
The second panel provides detailed information on the repository, notebooks, cells, modules, and their execution.
It displays the list of python modules used in each notebook to help users to see which modules are commonly used in a repository.
In the \textit{Execution Overview} tab, it provides information on which cell in the notebook resulted in the difference in outputs.
In case of failure in the execution of a notebook, it shows the detailed message on the reason of the occurring exception.
ReproduceMeGit provides an export feature supported by ProvBook to capture the prospective and retrospective provenance in RDF described by the REPRODUCE-ME ontology~\cite{sheebathesis}.
\section{Demonstration}
In our demonstration, users will be able to enter the URL of any GitHub repository containing Jupyter Notebooks and reproduce them.
They will be able to explore the results of ReproduceMeGit including the overview of (1) the GitHub Repository and the Jupyter Notebooks available in it, (2) the reproducibility study, (3) the notebooks which had same or different results, (4) the provenance of the execution of each notebook, (5) the modules used by each notebook, (6) the feature to export each notebook along with its provenance in RDF, and (7) the provenance history and provenance difference using ProvBook.
The source code and a demo video of ReproduceMeGit with an example are available at "\url{https://github.com/fusion-jena/ReproduceMeGit}".
\paragraph{Acknowledgments}\mbox{}\\
The authors thank the Carl Zeiss Foundation for the financial support of the project ``A Virtual Werkstatt for Digitization in the Sciences (K3)'' within the scope of the program-line ``Breakthroughs: Exploring Intelligent Systems'' for ``Digitization -- explore the basics, use applications''.


\bibliographystyle{unsrt}
\bibliography{demo}

\end{document}